\documentclass{article}
\usepackage{spconf,amsfonts,amssymb,amsmath,graphicx}

\newcommand{\R}{\mathbb{R}}

\title{Optimal Window and Lattice in Gabor Transform\\ Application to Audio Analysis}
\name{H. Lachambre $^1$\thanks{This research has been supported by EU FET Open grant UNLocX (255931).}, B. Ricaud $^2$, G. Stempfel $^1$, B. Torr\'esani $^3$, C. Wiesmeyr $^4$, D. M. Onchis $^5$}
\address{$^1$Genesis,	Domaine du Petit Arbois, 13545 Aix-en-Provence, France\\
helene.lachambre@genesis.fr, guillaume.stempfel@genesis.fr\\
$^2$Signal Processing Laboratory 2,	Ecole Polytechnique F\'ed\'erale de Lausanne (EPFL), \\Station 11, 1015 Lausanne, Switzerland - 
benjamin.ricaud@epfl.ch\\
$^3$Aix-Marseille Universit\'e, CNRS, Centrale Marseille, LATP, UMR 7353, 13453 Marseille, France\\
bruno.torresani@univ-amu.fr\\
$^4$Faculty of Mathematics, University of Vienna, Oskar-Morgenstern-Platz 1, 1090 Wien, Austria\\
christoph.wiesmeyr@univie.ac.at, darian.onchis@univie.ac.at\\
$^5$West University of Timisoara, Blvd. V. Parvan 4, Timisoara, 300223, Romania\\
donchis@hpc.uvt.ro}

\date{}

\begin{document}
\maketitle
\begin{abstract}
This article deals with the use of optimal lattice and optimal window in Discrete Gabor Transform computation. In the case of a generalized Gaussian window, extending earlier contributions, we introduce an additional local window adaptation technique for non-stationary signals. We illustrate our approach and the earlier one by addressing three time-frequency analysis problems to show the improvements achieved by the use of optimal lattice and window: close frequencies distinction, frequency estimation and SNR estimation. The results are presented, when possible, with real world audio signals.
\end{abstract}
\section{Introduction}
\label{sec:intro}
Our work takes place in the general field of Time-Frequency (T-F) signal analysis~\cite{Flandrin99}, in which signals are represented by a time-frequency image, obtained through a linear transform. We consider here the very well known and often used Discrete Gabor Transform (DGT), with a generalized Gaussian analysis window. For an efficient T-F signal analysis, all the parameters of the transform should be appropriately chosen, such as the window shape or length~\cite{Boashash03}. In a prior work~\cite{Jaillet07time}, a study was ran on the search for the optimal real Gaussian window. The optimization criterion was the minimization of the Renyi entropy~\cite{Renyi60} or of the Shannon entropy.

In a recent study~\cite{Ricaud13}, our main focus was to define and compute an optimal generalized Gaussian window for a given signal. The optimization criterion is the maximization of the energy concentration in the analysis domain, measured using $l_p$ norm. An optimization of the lattice parameters was also proposed in the same article. In the present article, we propose to demonstrate, on real world audio signals, the improvements these two methods actually provide on three DGT based analysis tasks: close frequencies distinction, frequency estimation and SNR estimation.

In the remaining of this article, we will briefly recall the main notations and the methods for the optimization of the window and of the lattice in Section~\ref{sec:theory}. Then, in Section~\ref{sec:applications}, we present some applications with real world audio signal.

\section{Theoretical background}
\label{sec:theory}
The theoretical setting of the window optimization principle and algorithms has been submitted separately~\cite{Ricaud13}. For the sake of completeness, we will only remind the main points. We work in the setting of finite dimensional signal spaces $\mathbb{C}^N$; let $g\in\mathbb{C}^N$ denote an analysis window, and $\Lambda\subset\mathbb{Z}^N\times\mathbb{Z}^N$ denote a T-F lattice. The DGT of a signal $f\in\mathbb{C}^N$ is defined as ${\cal V}_{g}f(x,\xi) = \sum_{n = 0}^{N}{f(n)g(n-x)e^{-j2\pi \xi n/N}}$ with $(x,\xi)\in\Lambda$ a point in the lattice.
\subsection{Window optimization}
\label{ssec:windowOpti}
In this part, our aim is to find the optimal apodization window for the energy concentration of the DGT. Therefore, as proposed in~\cite{Jaillet07time}, we quantify the quality of a T-F representation by measuring its energy concentration. For this purpose, we consider the $l_p$ norm of the DGT. $l_p$ norm is a measure of spreading for $p<2$ and of sparsity for $p>2$. As we want to optimize sparsity, we choose to maximize $l_p$ norm with $p>2$ ($p = 2.5$), since this leads to differentiable criteria~\cite{Ricaud13}.

The problem we solve may be written as:

\begin{equation}\label{optprob}
\mathop{\rm Argmax}_{g,\|g\|_2=1} \| {\cal V}_{g}f\|_p^p=\mathop{\rm Argmax}_{g,\|g\|_2=1}  \sum_{\xi,x\in\Lambda}|{\cal V}_{g}f(x,\xi) |^p,
\end{equation}

In order to be sure to obtain interpretable windows, we restrict the search to the family of chirped Gaussian. These functions may be written as:
\begin{equation}\label{ParamGaussian}
g_{\sigma,s}(t)=\sqrt[4]{\frac{2}{N\sigma}}e^{-\pi\frac{t^2}{N\sigma}+i \pi s t^2\frac{N+1}{N}},
\end{equation}
where $\sigma > 0$ is the spreading in time and $s\in\R$ is the chirping parameter and $N$ the signal length. The optimization problem becomes:
\begin{equation}\label{optprobparam}
\mathop{\rm Argmax}_{\sigma,s} \| {\cal V}_{g_{\sigma,s}}f\|_p^p.
\end{equation}

\begin{figure}[htb]
  \centering
  \centerline{\includegraphics[width=0.5\linewidth]{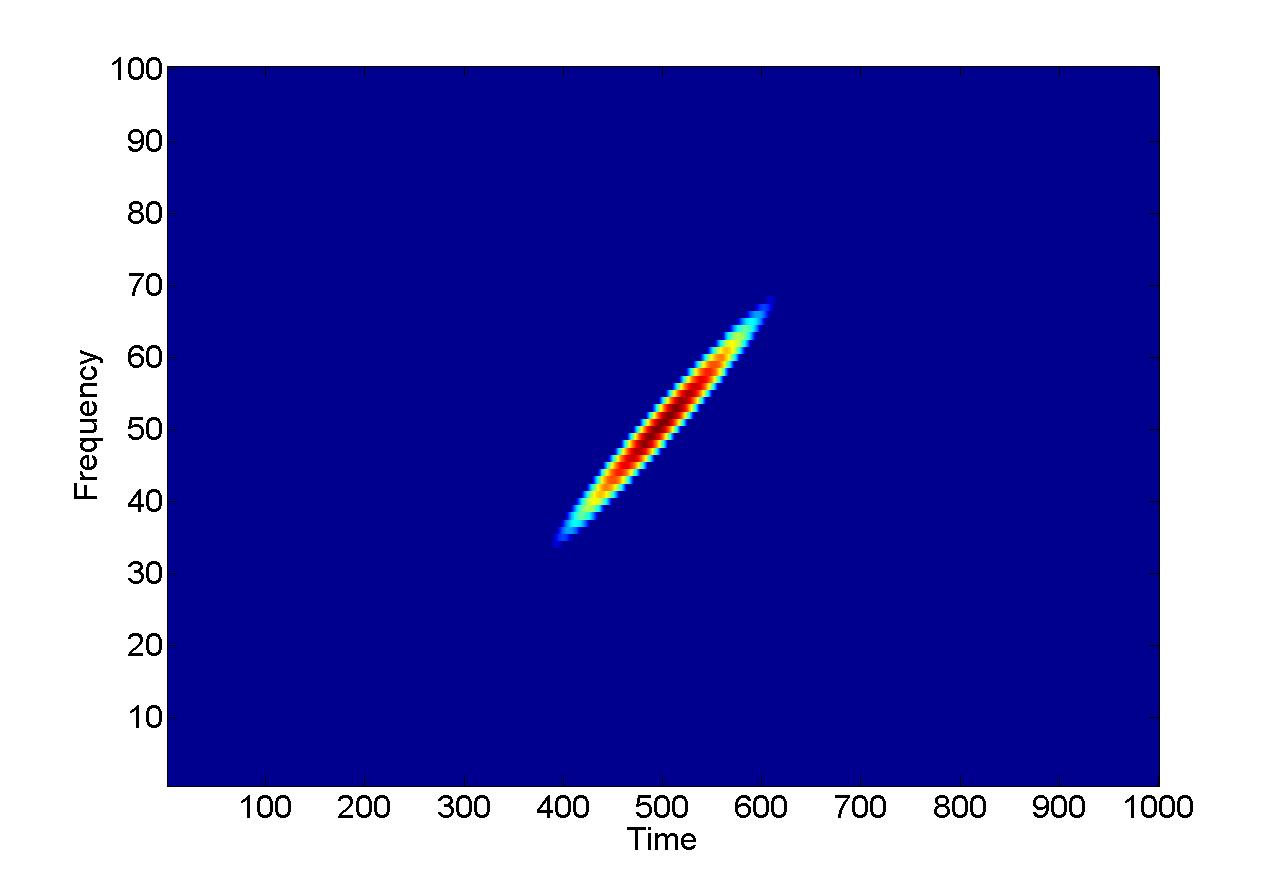}}
\caption{Ambiguity function of a chirped Gaussian.}
\label{fig:ChirpedGaussian}
\end{figure}

An example of the ambiguity function ${\cal V}_gg$ (the DGT of the window, analyzed by itself) of such a window is depicted on Fig.~\ref{fig:ChirpedGaussian}

\subsection{Lattice optimization}
The second problem we have is to find, for a given redundancy $R = \left| \Lambda\right|/N$, the best lattice, i.e. the optimal placement of the time frequency atoms.

A well chosen lattice is defined as a lattice that leads to a numerically stable inverse Gabor transform. The optimal lattice will therefore be the one that leads to an inverse Gabor transform as stable as possible.

Under the assumption that the hexagonal lattice is the optimal one for a standard Gaussian~\cite{Wiesmeyr13}, the optimal lattice for the optimal window $g_{\sigma,s}$ is~\cite{Ricaud13}:
\begin{equation}
\Lambda = \sqrt{\frac{N}{R}}
\left(\begin{array}{cc}
	1 & 0\\
	s & 1
\end{array}\right)
\left(\begin{array}{cc}
	\sqrt{\sigma} & 0\\
	0 & 1/\sqrt{\sigma}
	\end{array}\right)
\left(\begin{array}{cc}
	\frac{\sqrt[4]{3}}{\sqrt{2}} & 0\\
	\frac{1}{\sqrt[4]{3}\sqrt{2}} & \frac{\sqrt{2}}{\sqrt[4]{3}}
\end{array}\right)
\mathbb{Z}_N^2
\label{eq:optimallattice}
\end{equation}
where the operation matrices from left to right are the shearing, dilation and hexagonal sampling matrices. The constant in front gives the correct redundancy.

\subsection{Local window optimization and its evolution over time}
As may be seen on Fig.~\ref{fig:aircraft}.a and Fig.~\ref{fig:aircraft}.b, a unique analysis window may not be adapted to the whole signal, but only to a limited T-F region. In this work, we time-locally adapt the analysis window, with a constant lattice over the whole signal. Considering that in real world signals, the evolution of the amplitudes and frequencies are smooth, we look for a smooth evolution of the window over time. More precisely, we proceed as follows:
\begin{itemize}
\item split the signal in $M$ parts,
\item on each part, find the optimal window as presented in part~\ref{ssec:windowOpti},
\item consider this window optimal on the middle frame of the part, and interpolate the $\sigma$ and $s$ parameters over every frame,
\item compute the DGT, considering a different window on each frame.
\end{itemize}

Note that the so-computed DGT is in fact a non-stationnary Gabor Transform~\cite{Balazs11theory}, and that a careful exploration of its invertibility should be explored.

Note also that the first step (choice of $M$ and position of boundaries) is, at this time, fully manually supervised.

\label{ssec:local}

\section{Applications}
\label{sec:applications}
We propose to evaluate the usefulness of this method on three applications:
\begin{itemize}
\item very close frequencies distinction,
\item frequency estimation of the main harmonic,
\item signal-to-noise ratio estimation.
\end{itemize}

Our aim being to present the improvement achieved by use of optimally chirped Gaussian windows, we consider here applications to real signals.

For each application, we propose a naive approach based on the analysis of the DGT. We are then able to compare the performances obtained with a DGT computed with different lattices and with different windows. The optimal lattice will be compared to other lattices with the same redundancy. The optimally chirped and dilated Gaussian window will be compared to an optimally dilated real valued Gaussian window. This last window is estimated with a process similar to equation~\ref{optprobparam}, with one parameter:

\begin{equation}
\mathop{\rm Argmax}_{\sigma} \| {\cal V}_{g_{\sigma}}(f)\|_p^p.
\end{equation}

\subsection{Close frequencies distinction}
\label{ssec:closefreq}

\begin{figure}[h!]
\begin{minipage}[b]{\linewidth}
  \centering
  \centerline{\includegraphics[width=.8\linewidth]{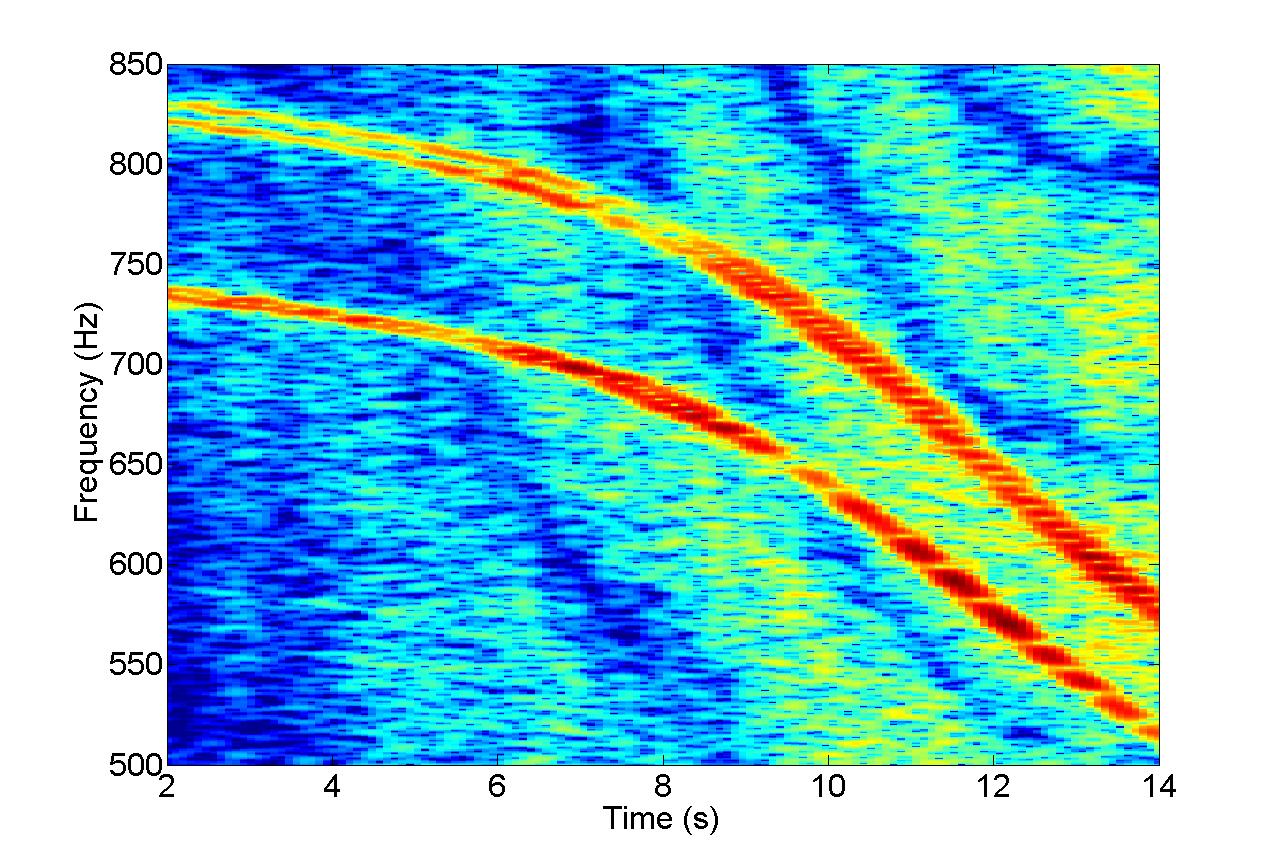}}
  \centerline{(a) Optimal real Gaussian}\medskip
\end{minipage}
\hfill
\begin{minipage}[b]{\linewidth}
  \centering
  \centerline{\includegraphics[width=0.8\linewidth]{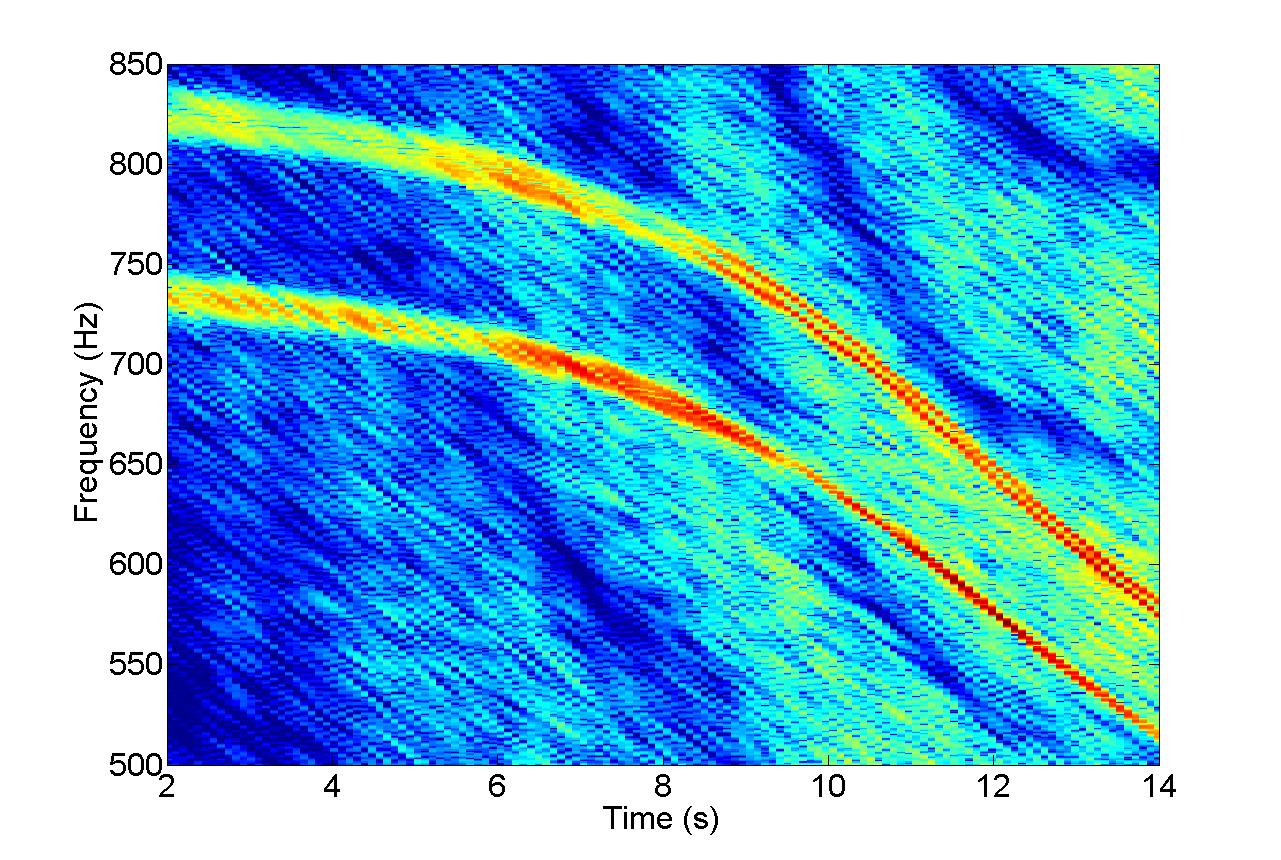}}
  \centerline{(b) Optimal chirped Gaussian}\medskip
\end{minipage}
\hfill
\begin{minipage}[b]{\linewidth}
  \centering
  \centerline{\includegraphics[width=0.8\linewidth]{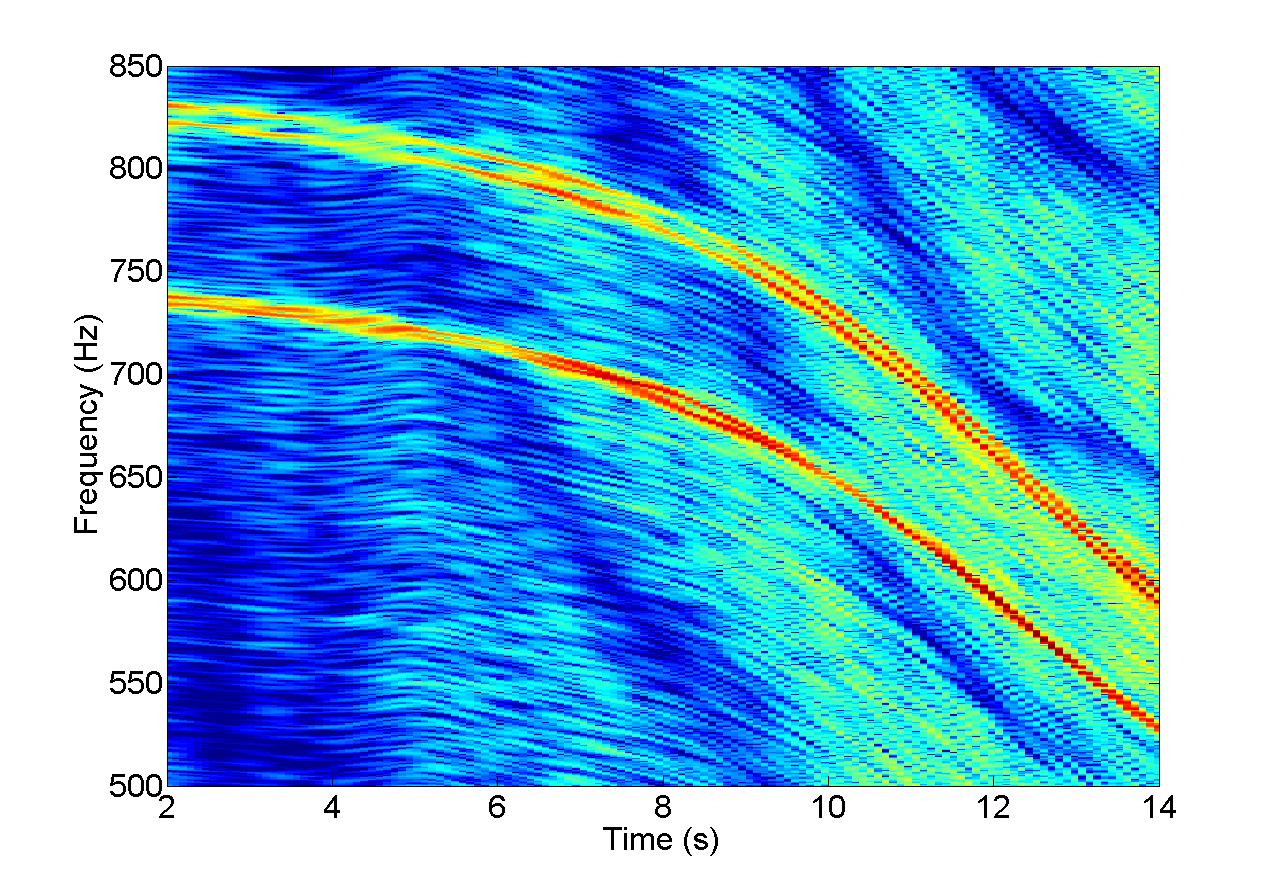}}
  \centerline{(c) Locally optimal chirped Gaussian}\medskip
\end{minipage}
\caption{Aircraft passing. On figure b, one can distinguish very close frequencies on the second part of the signal. On figure c, one can distinguish them all along the signal.}
\label{fig:aircraft}
\end{figure}

\begin{figure}[h!]
\begin{minipage}[b]{\linewidth}
  \centering
  \centerline{\includegraphics[width=0.8\linewidth]{3_Aircraft-Passing-OptimalWin.jpg}}
  \centerline{(a) Optimal lattice}\medskip
\end{minipage}
\hfill
\begin{minipage}[b]{\linewidth}
  \centering
  \centerline{\includegraphics[width=.8\linewidth]{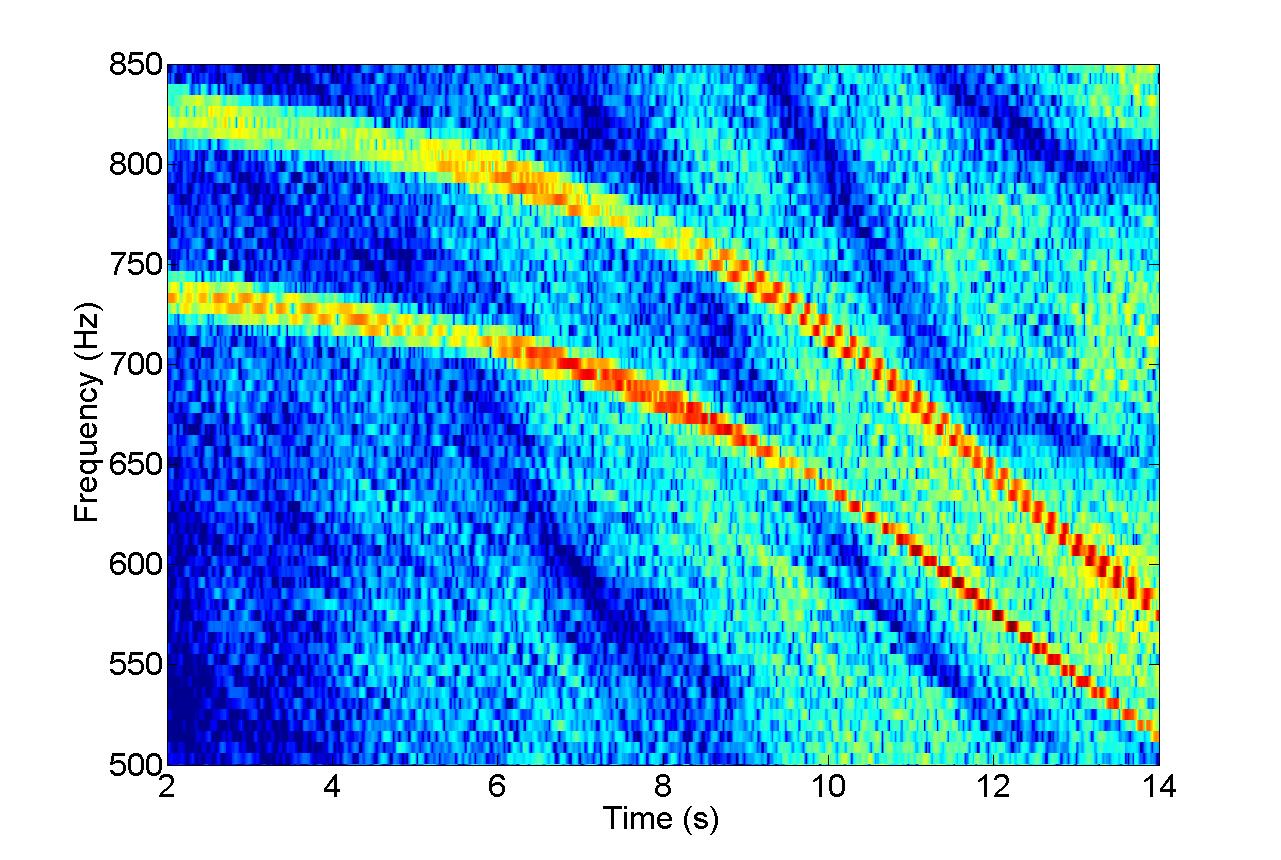}}
  \centerline{(b) More time bins, less frequency bins}\medskip
\end{minipage}
\hfill
\begin{minipage}[b]{\linewidth}
  \centering
  \centerline{\includegraphics[width=0.8\linewidth]{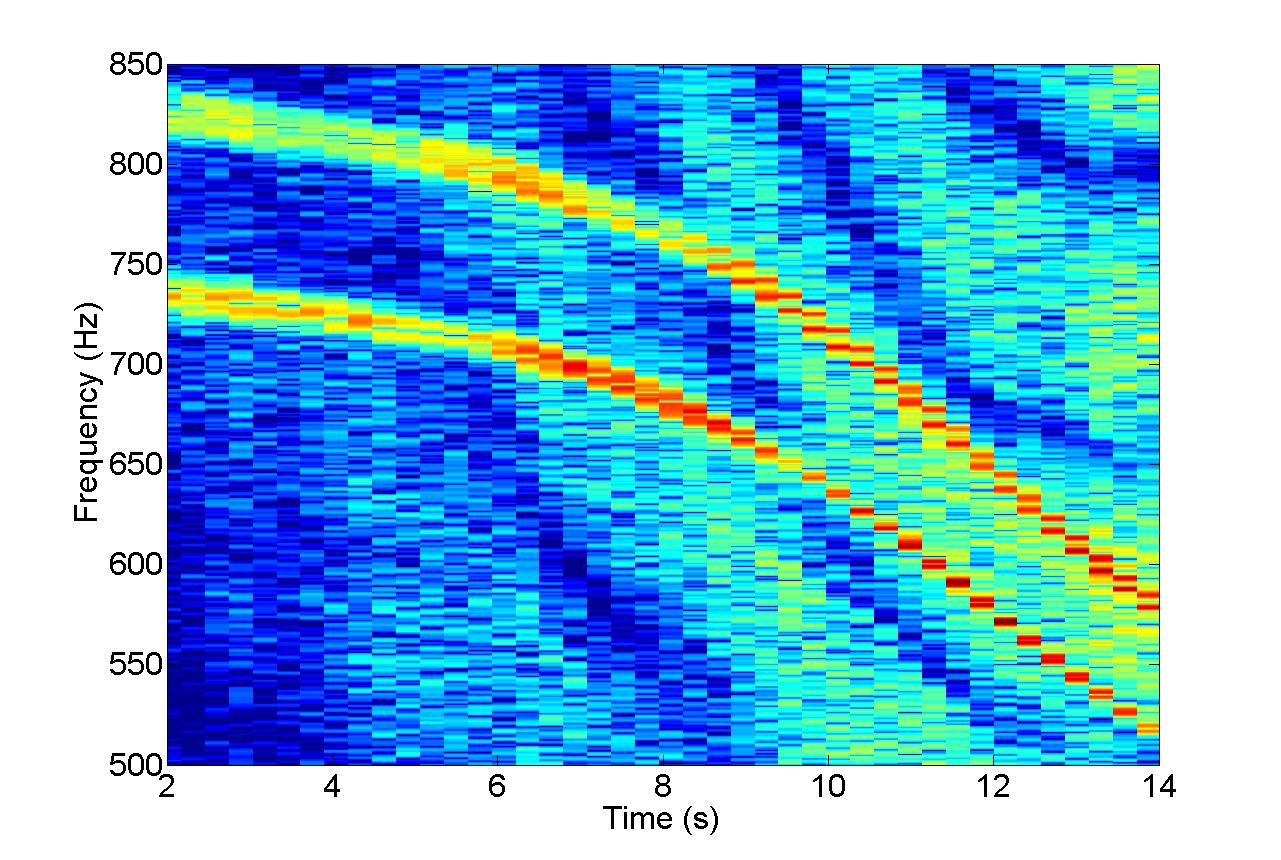}}
  \centerline{(c) Less time bins, more frequency bins}\medskip
\end{minipage}
\caption{Sound of an aircraft passing (DGT) using the same windows and redundancies but different lattices. The optimal lattice (a) shows a better behavior of the DGT.}
\label{fig:aircraftlattice}
\end{figure}

In this application, our proposition is mainly visually validated. The example presented in Fig.~\ref{fig:aircraft} and Fig.~\ref{fig:aircraftlattice} is the recording of an aircraft passing. The aircraft produces several different frequencies, some of them are very close. Furthermore, these frequencies evolve over time due to the Doppler shift. On Fig.~\ref{fig:aircraft}.a, the signal is analyzed with a optimally dilated real valued  Gaussian, which is adapted for the beginning of the signal, while on Fig.~\ref{fig:aircraft}.b, the window is the globally optimal chirped Gaussian, which is adapted for the end of the signal. On Fig.~\ref{fig:aircraft}.c, the window is locally adapted, with a splitting of the signal into 25 1-second parts, the close frequencies are visible all along the signal. All these DGTs are computed with the same lattice.

Fig.~\ref{fig:aircraftlattice} allows one to visualize the improvement provided by the use of the optimal lattice. The same signal is analyzed with the same optimal window and the same redundancy, with the optimal lattice (Fig.~\ref{fig:aircraftlattice}.a), and with two non-optimal lattices: more time bins and less frequency bins (Fig.~\ref{fig:aircraftlattice}.b), or the contrary (Fig.~\ref{fig:aircraftlattice}.c). 

\subsection{Frequency estimation}
\label{ssec:freqest}
\begin{figure}[h!]
\begin{minipage}[b]{\linewidth}
  \centering
  \centerline{\includegraphics[width=.8\linewidth]{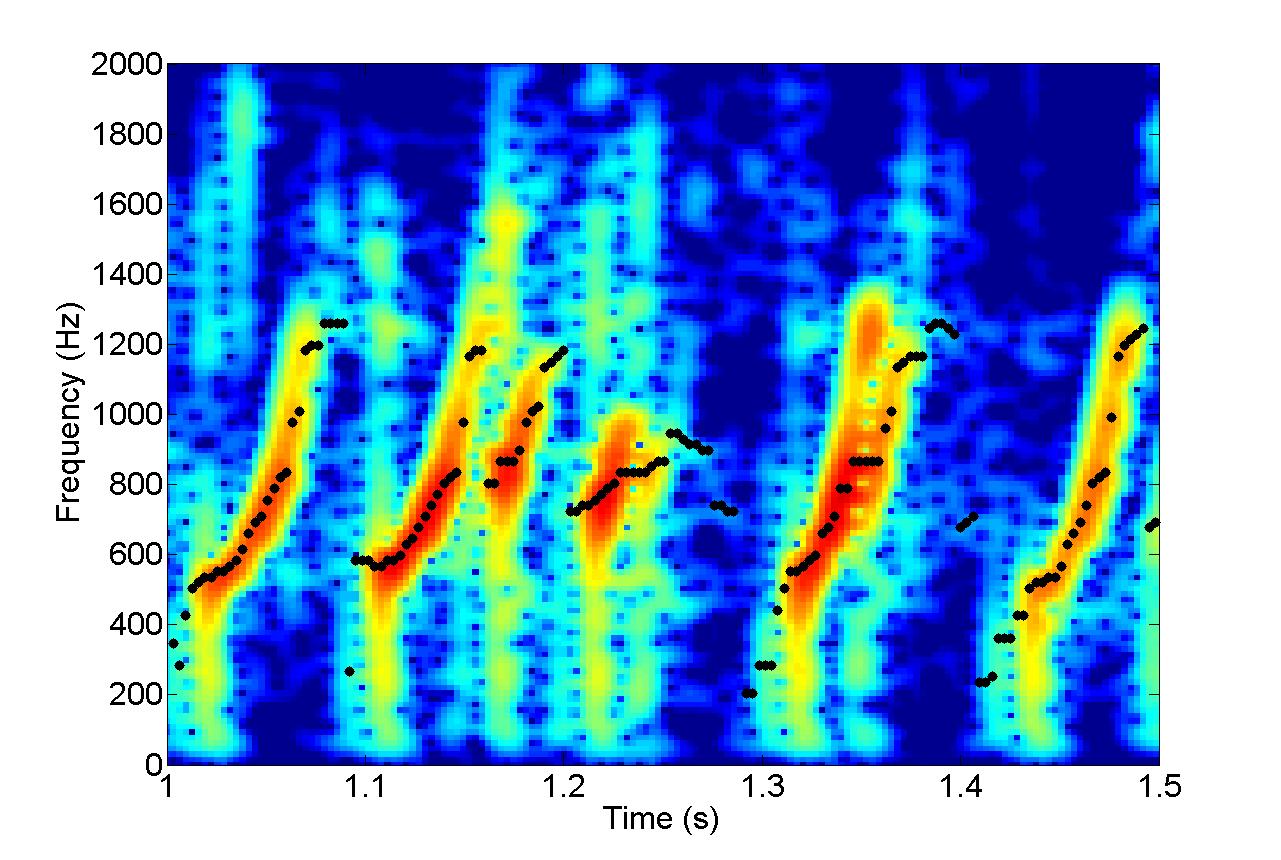}}
  \centerline{(a) Optimal real Gaussian}\medskip
\end{minipage}
\hfill
\begin{minipage}[b]{\linewidth}
  \centering
  \centerline{\includegraphics[width=0.8\linewidth]{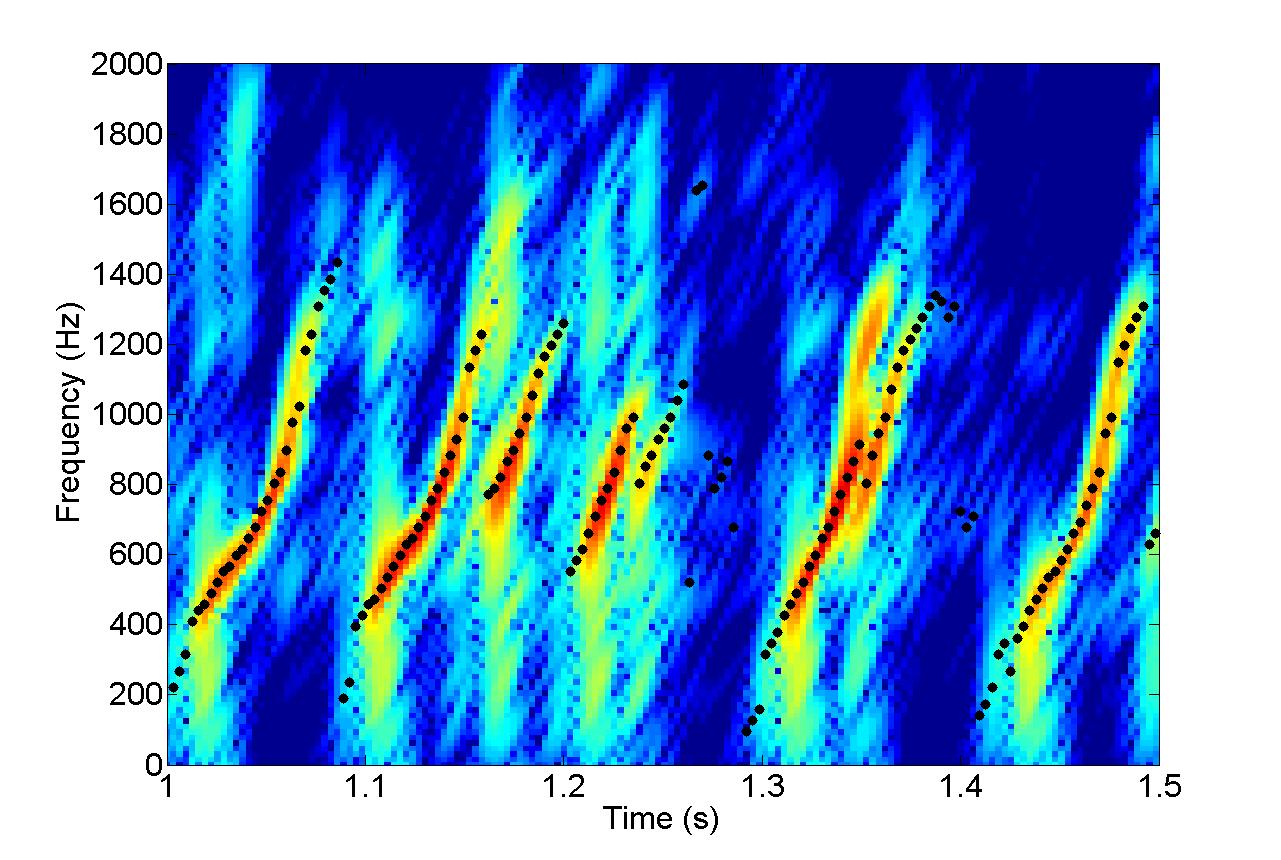}}
  \centerline{(b) Optimal chirped Gaussian}\medskip
\end{minipage}
\caption{Bubbles in water. The estimated main frequency is highlighted with dark points all along the DGT.}
\label{fig:bubbles}
\end{figure}

In this application/example, we suppose that we are tracking a single frequency on each frame. Therefore, we propose to pick up, on each frame, the frequency corresponding to the maximum amplitude. On the example presented on Fig.~\ref{fig:bubbles}, the signal is composed of several bubbles bursting in water. The sound of each bursting bubble leads to a clearly recongnizable pattern in the TF plane. On Fig.~\ref{fig:bubbles}.a, the DGT is computed with an optimal non-chirped Gaussian window, while on Fig.~\ref{fig:bubbles}.b, the window is an optimally chirped Gaussian. On each figure, the estimated frequency is highlighted with black dots. On Fig.~\ref{fig:bubbles}.a, contrary to what was expected, the patterns are neither similar nor regular. On the other hand, on Fig.~\ref{fig:bubbles}.b the patterns are regular and similar. Since the frequency patterns do not change much along time, a local adaptation of the window would not provide much improvements. The optimal window reveals the acoustic signature of the bubbles, and could therefore allow further analysis of the signal: descriptive statistics on the signatures could provide physical information on the bubbles, such as their number, their size, or their velocity in water.

\subsection{SNR estimation}
\label{ssec:snrest}
In this application, we propose to estimate the Signal to Noise Ratio (SNR) of a noisy signal. In our example, the signal is a square chirp whose amplitude is log-linearly decreasing, the noise is a constant white Gaussian noise.

As the estimation of the noise power does not depend on the window (as long as the latter has a unit norm, which is true in our case), we may estimate it straightaway as the mean value of the non-harmonic part of the spectrum.

The amplitude of the square chirp is directly proportional to the amplitude of the maximum of the spectrum. The normalization factor due to the window is estimated with a known signal (pure sinus with known amplitude).

\begin{figure}[h!]
  \centering
  \centerline{\includegraphics[width=0.8\linewidth]{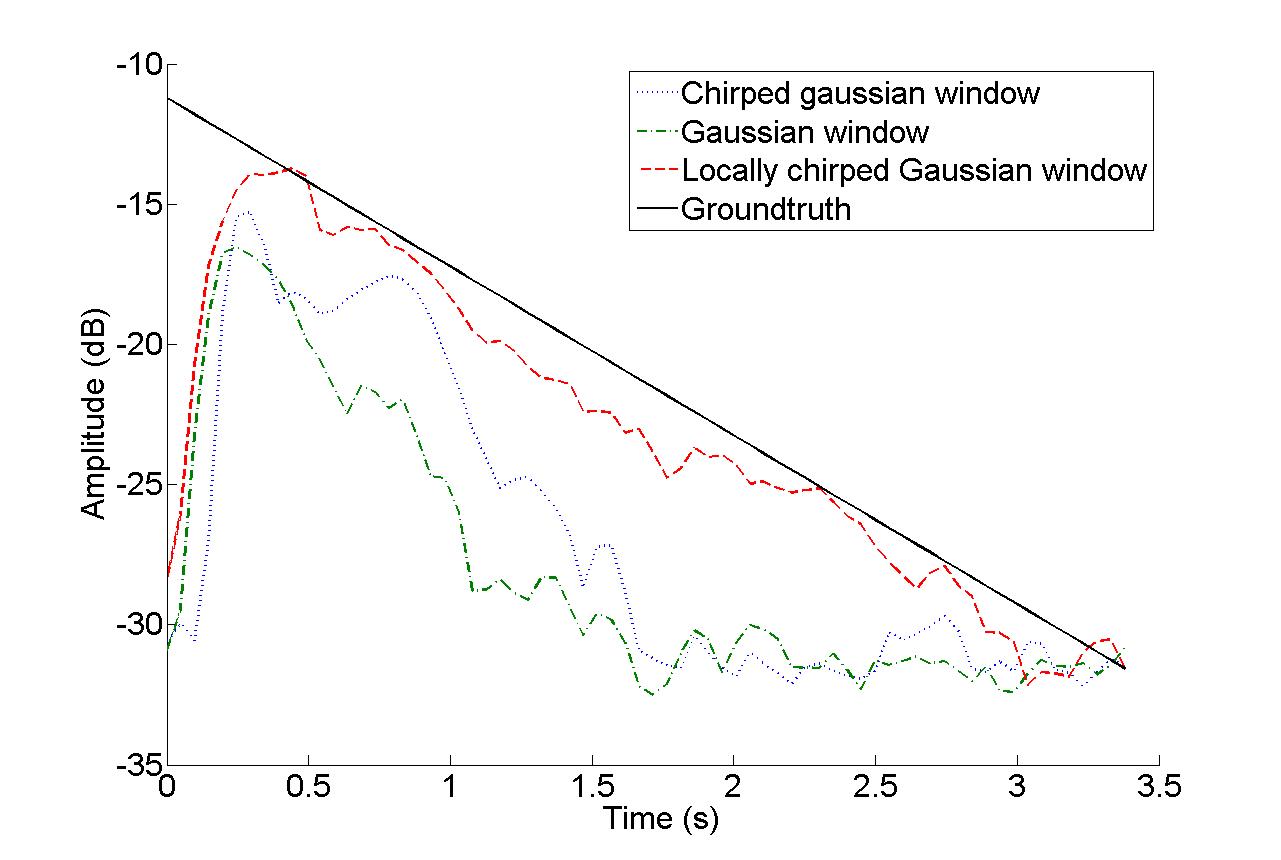}}
\caption{Square chirp (log-linearly decreasing amplitude) with constant white Gaussian noise. SNR estimated on DGTs computed with diverse windows.}
\label{fig:snr}
\end{figure}

With these two values available (instantaneous chirp power and noise power), we are able to compute the instantaneous SNR. The SNR values computed using various different windows, as well as the true SNR are plotted on Fig.~\ref{fig:snr}. The figure shows clearly that the best estimate is obtained with the optimal window, as could be anticipated.

\section{Conclusion and perspectives}
\label{sec:conclusion}
In this article, we presented the results of an optimization procedure that provides an optimal window and an optimal lattice for the computation of the DGT for some given signal.
Three different audio applications showcase the usefulness of the proposed procedure: distinction of close frequencies, frequency estimation and SNR estimation. For the first two applications, quantitative results could be estimated on real world audio signal, while a synthetic signal was used for the last application. All the examples show that the use of the optimal lattice and of the optimal window for the DGT computation leads to more interpretable and usefull results. 

Preliminary results were also presented, with the use of locally optimal windows, which are very promising. As this implies the use of non-stationnary Gabor Transform, the invertibility of such a DGT will have to be carefully investigated, along with the automatized search of the windows.

\section{Acknowledgements}
This research has been supported by EU FET Open grant UNLocX (255931). The work of last author of the article has been  partially supported by the EU grant no. FP7-REGPOT-CT-2011-284595 (HOST).

\bibliographystyle{IEEEbib}
\bibliography{refs}

\end{document}